

\def\NI{\noindent}
\long\def\UN#1{$\underline{{\vphantom{\hbox{#1}}}\smash{\hbox{#1}}}$}
\magnification=1215
\overfullrule=0pt
\hfuzz=16pt
\voffset=0.0 true in
\vsize=8.8 true in
\def\NP{\vfil\eject}
\baselineskip 20pt
\parskip 6pt
\hoffset=0.1 true in
\hsize=6.3 true in
\nopagenumbers
\pageno=1
\footline={\hfil -- {\folio} -- \hfil}

\hphantom{AA}

\hphantom{AA}

\centerline{\UN{\bf Annihilation of Immobile Reactants on
the Bethe Lattice}}

\vskip 0.4in

\centerline{\bf Satya~N.~Majumdar$^a$\ {\rm and}\
Vladimir~Privman$^b$}

\vskip 0.2in

\NI $^a$AT\&T Bell Laboratories, 600 Mountain Avenue, Murray Hill,
New Jersey 07974

\NI $^b$Department of Physics, Clarkson
University, Potsdam, New York 13699--5820

\vskip 0.6in

\centerline{\bf ABSTRACT}

Two-particle annihilation reaction, $A+A \to {\rm inert}$,
for immobile reactants on the Bethe lattice is solved exactly
for the initially random distribution. The process
reaches an absorbing state in which no nearest-neighbor
reactants are left. The approach of the concentration to the
limiting value is exponential. The solution reproduces the
known one-dimensional result which is further extended to
the reaction $A+B \to {\rm inert}$.

\vfill

\NI {\bf PACS numbers:}$\;$  82.20.Mj {\it and\/} 05.40.+j

\hphantom{AA}

\hphantom{AA}

\NP

Recent studies of reaction-diffusion systems
have emphasized fluctuation effects and breakdown of the
standard chemical rate equations in low dimensions.
For the simplest reactions of
two-particle coagulation, $A+A \to A$, and annihilation,
$A+A \to {\rm inert}$, on the one-dimensional lattice,
several exact results have been reported [1-13].
In the diffusion-limited, instantaneous-reaction case, these
processes show non-mean-field power-law decay of the
$A$-particle density.

Another solvable limit, in one dimension [14], is the
case of no diffusion at all. Generally such models of \UN{\sl
immobile reactants}\  have received less attention
in the recent literature [14-16]. The reason is that unless
longer-range reactions are allowed for [15-16], the time
dependence involves exponential relaxation to an absorbing
state rather than power-law behavior typical of the
fast-diffusion reactions. Thus there are no universal
fluctuation effects involved.

On the other hand, immobile-reactant systems provide an
example of freezing in an absorbing state with an explicitly
nonuniversal, initial-condition-dependent behavior
persistent at all times and, again, not consistent with the
mean-field rate equations. It is therefore of interest to
derive exact results whenever possible. Thus far, they were
available only in one dimension [14]. In this work we
report an exact solution for $A+A \to {\rm inert}$ on the
Bethe lattice. We also derive exact results for the
reaction $A+B \to {\rm inert}$, limited to one dimension.

Examination of the one-dimensional solution of $A+A \to {\rm
inert}$, [14], suggests close similarity to the models of
random sequential adsorption [17-19]. Specifically, the
annihilation reaction resulting in
removal of two nearest-neighbor fixed reactants is
equivalent to ``deposition'' of a ``dimer'' of two empty
(reacted) sites. The two processes are dual to each other in
that nearest-neighbor pairs of sites available for deposition
correspond to unreacted pairs of
reactants. Mathematically, the only difference is in the
initial conditions. In deposition, the lattice is usually
assumed empty at time $t=0$ which would correspond to the
full occupancy for reaction.

This connection to random sequential adsorption
models suggests that exact solutions can be sought for
models of multiparticle annihilation reactions
corresponding to $n$-mer deposition [19], and for models
formulated on high-connectivity lattices such as the Bethe
lattice, etc.; see [18,20-21]. In this work we consider the
Bethe-lattice case; the methods of [20-21] are adapted for
reactions. Besides notational differences, this
essentially amounts to a more careful treatment of random
initial conditions. The latter is achieved by a method
different from the techniques used in deposition model
studies [20-23]. Since the Bethe lattice of coordination
number 2 is identical to the one-dimensional lattice, we
obtain the $d=1$ solution as well. However, further
simplification in $d=1$ allows us to also solve exactly the
reaction $A+B \to {\rm inert}$.

The Bethe lattice of coordination number $z \geq 2$ can be
viewed as an interior part of the infinite Cayley tree: each
site is connected by bonds to $z$ nearest neighbor sites, and
there are no closed loops formed by bonds. In fact the
details of the lattice connectivity are not important for
our considerations. However, we disregard any end-effects.
Since the number of ``boundary'' sites in a
finite-number-of-generations Cayley tree grows
proportional to the total number of sites provided $z>2$
(i.e., there is branching in each generation), the boundary
effects can be profound when long-range spatial correlations
are present such as at phase transitions, both static [24]
and dynamical [25]. The model considered here does
not have any ``dangerous'' spatial correlations; size effects
can be safely ignored.

Consider a $k$-site connected cluster on the Bethe lattice.
One interesting feature of a loopless lattice, shared with
the $d=1$ lattice for which $z=2$, is that in such a cluster
the $k$ sites are connected by exactly $k-1$ internal bonds.
This statement is well known and easily established by
induction: each new site can only be connected to one
existing cluster site, by one bond, because loops are not
possible. Another useful conclusion is that the number of
bonds shared by the cluster sites and the nearest neighbor
sites immediately outside the cluster under consideration, is
$zk-2(k-1)$. Here $zk$ is the total number of neighbors
seen by all the $k$ cluster sites, while the term $2(k-1)$
subtracts the number of neighbors internal to the cluster
(twice the number of bonds).

In the model of immobile reactants we assume that the
lattice sites are initially occupied at random with
probability $\rho$, and empty with probability $1-\rho$.
The initial reactant density per site is $c(0)=\rho$,
and we would like to calculate the time-dependence of the
density $c(t)$ at later times $t>0$, given that each
nearest-neighbor reactant pair annihilates with the rate
$R$ per unit time. In fact, it is convenient to absorb the
rate in the dimensionless time variable

$$ \tau=Rt \;\; . \eqno(1) $$

Similar to the random sequential adsorption studies
[17,22] and some recent results for particle-exchange
dynamical models [26-27], we consider the probability $P_k
(\tau ) $ that a connected $k$-site cluster is fully occupied
by reactants at time $t$. Configuration of the sites which
are nearest-neighbor but exterior to the cluster can be
arbitrary. Thus, initially,

$$ P_k(0)=\rho^k \;\; . \eqno(2) $$

At times $\tau > 0$, the quantities $P_k(\tau)$ remain
the same for all cluster topologies on a loopless lattice,
provided the initial conditions are topology independent.
This is because the topology dependence is not generated
dynamically by the evolution equations for the $P_k(\tau)$.
Indeed, their time variation is only determined by the number
of possible reaction events within the cluster, which
is equal the number of bonds, $k-1$, and the number of
possible reaction events in which the reacting pair has one
site within the cluster and another outside the cluster. The
latter again is not dependent on the cluster topology for
loopless clusters. We argued earlier that the number of such
pairs is $zk-2k+2$. Thus, the time-dependence can be
obtained from the relations

$$ {dP_k \over d\tau} = -(k-1)P_k -(zk-2k+2)P_{k+1} \;\; .
\eqno(3) $$

\NI Here the first term is self-explanatory. In the second
term, the probability $P_{k+1}$ is used because a larger
cluster must be actually occupied in order for a reaction
event involving a site outside the original $k$-site cluster
to proceed.

Relations (3) apply for all $k \geq 1$. The most
interesting quantity is $P_1(\tau ) = c(\tau ) $. It is
expected to decrease in time but remain finite as $\tau \to
\infty$. All other probabilities $P_{k>1}$ are expected to
vanish for large times.

The solution of the recursion relations (3) can be obtained
by various methods. Perhaps the simplest is to note that
the Ansatz

$$ P_k(\tau )=c(\tau ) \left[\sigma(\tau )\right]^{k-1} \;\;
, \eqno(4)$$

\NI where $\sigma(0)=\rho $, eliminates the $k$-dependence.
Indeed, substitution in (3) shows that the solution of the
form (4) is possible provided

$$ {d\sigma \over d\tau } = -\sigma-(z-2)\sigma^2 \;\; ,
\eqno(5) $$

$$ {dc \over d\tau } = -z\sigma c \;\; . \eqno(6) $$

The solution is

$$ \sigma={\rho e^{-\tau}  \over 1+(z-2)\rho
\left(1-e^{-\tau} \right) } \;\; , \eqno(7) $$

$$ c=\rho \left[ 1+(z-2)\rho \left(1-e^{-\tau}
\right) \right]^{-z/(z-2)} \;\; . \eqno(8) $$

\NI For $z=2$ the $d=1$ solution is obtained either as a
limit of (8) or directly,

$$ c_{z=2}= \rho e^{-2\rho \left(1-e^{-\tau}
\right) } \;\; . \eqno(9) $$

\NI An interesting feature of expressions (8)-(9) is the
explicit nonlinear dependence of the surviving reactant
density for all times on the initial density $\rho$. The
approach to the limiting density as $\tau \to
\infty$ is exponential, $\sim e^{-\tau}\;$.

We now turn to the strictly one-dimensional case.
Additional restrictions on the cluster topology now allow
solution of the two-species reaction $A+B \to {\rm inert}$.
Let us assume that initially the $d=1$ lattice
sites are occupied by reactant species $A$  with
probability $\alpha$, and by species $B$ with probability
$\beta$, where $\alpha+\beta \leq 1$. The
sites are empty with probability $1-\alpha-\beta$.
Nearest-neighbor $AB$ and $BA$ pairs react with rate $R\,$;
see (1).

As before, we consider the probability that a
$k$-site-long cluster is ``fully reactive.'' Let $A_k
(\tau )$ denote the fraction of $k$-site clusters which are
fully filled, with reactants in the configuration
$ABAB\ldots$, i.e., the leftmost site is $A$ and the
sequence is fully alternating. Similarly, let
$B_k (\tau) $ denote the fraction of $k$-site clusters of
the type $BABA\ldots$, with the leftmost site $B$ and
otherwise fully alternating order. The state of the
neighbor sites outside the $k$-cluster is not important in
the definition of these probabilities which are conditioned
only on the internal cluster configuration.

Let us introduce the quantities

$$ \omega = {\alpha + \beta \over 2} \;\; , \eqno(10) $$

$$ \rho=\sqrt{\alpha \beta} \;\; , \eqno(11) $$

\NI where the notation $\rho$ in (11) will become clear
later on. Then initially we have

$$ A_{\hbox{$k$-even}} (0) = B_{\hbox{$k$-even}} (0) = \rho^k
\;\; , \eqno(12) $$

$$ A_{\hbox{$k$-odd}} (0) /\alpha =  B_{\hbox{$k$-odd}} (0)
/ \beta = \rho^{k-1} \;\; . \eqno(13) $$

Let us denote the concentrations, per site, of reactant
species $A$ and $B$ by $a(\tau ) \equiv A_1 (\tau)$ and
$b(\tau ) \equiv B_1 (\tau)$, respectively. We have

$$ a(0) = \alpha \;\; , \eqno(14) $$

$$ b(0) = \beta \;\; , \eqno(15) $$

$$ a(\tau ) - b(\tau ) = \alpha - \beta \;\; . \eqno(16) $$

\NI The latter relation is obvious. Thus we only have to
calculate the sum $a+b$. It proves useful to introduce the
quantities

$$ P_k( \tau ) = \left[A_k (\tau ) +B_k (\tau ) \right]\big
/ 2 \;\; , \eqno(17) $$

\NI so that the sum $a+b$ is given by $2P_1$.

Now the probabilities $A_k$ and $B_k$ satisfy the relations

$$ {dA_k \over d\tau} = -(k-1)A_k -A_{k+1} -B_{k+1} \;\; ,
\eqno(18) $$

$$ {dB_k \over d\tau} = -(k-1)B_k -B_{k+1} -A_{k+1} \;\; .
\eqno(19) $$

\NI Here the first term corresponds to internal pairs
reacting, the second term corresponds to the rightmost site
reacting with an occupied external site, while the third
term in both relations corresponds to the leftmost site
reacting ``externally.''

For $P_k$ we get, by summing (18) and (19),

$$ {dP_k \over d\tau} = -(k-1)P_k -2P_{k+1} \;\; ,
\eqno(20) $$

\NI which is in fact identical to the $d=1$ random
sequential adsorption recursion [17,19,22], as well as to
the $z=2$ variant of (3). However, the initial conditions are
more complicated,

$$ P_{\hbox{$k$-even}} (0) = \rho^k
\;\; , \eqno(21) $$

$$ P_{\hbox{$k$-odd}} (0) = \omega \rho^{k-1}
\;\; . \eqno(22) $$

The solution is obtained by methods similar to solving the
Bethe-lattice recursions. We try forms similar to (4),

$$ P_{\hbox{$k$-even}}
=C_{\hbox{even}}(\tau)
\left[{\Sigma}(\tau) \right]^{k-1} \;\; ,
\eqno(23) $$

$$ P_{\hbox{$k$-odd}}
=C_{\hbox{odd}}(\tau)
\left[{\Sigma}(\tau) \right]^{k-1} \;\; .
\eqno(24) $$

\NI These, when substituted in (20), yield coupled
first-order differential equations which can be solved
explicitly to give

$$2P_k=\rho^{k-1} e^{-(k-1) \tau }
\left[(\rho+\omega) e^{-2\rho \left(1-e^{-\tau} \right)} +
(-1)^k (\rho-\omega) e^{2\rho \left(1-e^{-\tau} \right)}
\right] \;\; . \eqno(25) $$

\NI The reactant concentrations then follow as

$$ a(\tau)={1 \over 2}
\left[\alpha - \beta + (\rho+\omega) e^{-2\rho
\left(1-e^{-\tau} \right)} - (\rho-\omega) e^{2\rho
\left(1-e^{-\tau} \right)}  \right] \;\; , \eqno(26) $$

$$ b(\tau)={1 \over 2}
\left[ \beta - \alpha + (\rho+\omega) e^{-2\rho
\left(1-e^{-\tau} \right)} - (\rho-\omega) e^{2\rho
\left(1-e^{-\tau} \right)}  \right] \;\; , \eqno(27) $$

\NI where the parameters are related via (10)-(11).

As in the fast-diffusion case, which was not solved exactly
even in $d=1$ but only analyzed asymptotically [7,28], the
functional form of the concentration is different depending
on if the initial concentrations are equal or not, although
the difference here is less spectacular. For $\alpha =
\beta$\ $(=\omega=\rho)$, the time-dependence of \UN{\sl
each}\  of the species concentrations is identical to the
$d=1$ result (9). The time-dependence via the
double-exponential expressions entering the
general results (26) and (27) is also similar to (9)
with the effective concentration given by the geometrical
mean $\rho$; see (11). However, for $\alpha \neq \beta$ the
full time-dependence is more complicated than for the
single-species reaction, involving both the
double-exponential term and its inverse.

In summary we presented exact solutions for two-particle
annihilation reaction on the Bethe lattice. For the
coordination number $z>2$, results were reported for the
single-species case. In $d=1$, where $z=2$, we reproduced the
known exact single-species result and extended the solution
to two-species reaction.

\NP

\centerline{\bf REFERENCES}

{\frenchspacing

\item{[1]} M. Bramson and D. Griffeath,
Ann. Prob. {\bf 8}, 183 (1980).

\item{[2]} D.C. Torney and H.M. McConnell,
J. Phys. Chem. {\bf 87}, 1941 (1983).

\item{[3]} K. Kang, P. Meakin, J.H. Oh and S. Redner,
J. Phys. A{\bf 17}, L665 (1984).

\item{[4]} T. Liggett, {\sl Interacting Particle Systems\/}
(Springer-Verlag, New York, 1985).

\item{[5]} Z. Racz, Phys. Rev. Lett. {\bf 55}, 1707 (1985).

\item{[6]} A.A. Lushnikov, Phys. Lett. A{\bf 120}, 135 (1987).

\item{[7]} M. Bramson and J.L. Lebowitz, Phys. Rev. Lett. {\bf 61},
2397 (1988).

\item{[8]} C.R. Doering and D. ben--Avraham, Phys. Rev. A{\bf 38},
3035 (1988).

\item{[9]} V. Kuzovkov and E. Kotomin, Rep. Prog. Phys.
{\bf 51}, 1479 (1988).

\item{[10]} J.L. Spouge, Phys. Rev. Lett. {\bf 60},
871 (1988).

\item{[11]} J.G. Amar and F. Family, Phys. Rev. A{\bf 41}, 3258
(1990).

\item{[12]} D. ben--Avraham, M.A. Burschka and C.R. Doering,
J. Stat. Phys. {\bf 60}, 695 (1990).

\item{[13]} V. Privman, J. Stat. Phys. {\bf 69}, 629 (1992).

\item{[14]} V.M. Kenkre and H.M. Van Horn, Phys. Rev.
A{\bf 23}, 3200 (1981).

\item{[15]} H. Schn\"orer, V. Kuzovkov and A. Blumen,
Phys. Rev. Lett. {\bf 63}, 805 (1989).

\item{[16]} H. Schn\"orer, V. Kuzovkov and A. Blumen,
J. Chem. Phys. {\bf 92}, 2310 (1990).

\item{[17]} M.C. Bartelt and V. Privman, Int. J. Mod. Phys.
B{\bf 5}, 2883 (1991).

\item{[18]} J.W. Evans, {\sl Random and Cooperative
Sequential Adsorption}, Rev. Mod. Phys., in print (1993).

\item{[19]} J.J. Gonzalez, P.C. Hemmer  and J.S. H{\o}ye,
Chem. Phys. {\bf 3}, 228 (1974).

\item{[20]} J.W. Evans, J. Math. Phys. {\bf 25}, 2527
(1984).

\item{[21]} Y. Fan and J.K. Percus, Phys. Rev. A{\bf 44},
5099 (1991).

\item{[22]} E.R. Cohen and H. Reiss, J. Chem. Phys. {\bf
38}, 680 (1963).

\item{[23]} M.J. de Oliveira, T. Tom\'e and R. Dickman,
Phys. Rev. A{\bf 46}, 6294 (1992).

\item{[24]} M.L. Glasser and M.K. Goldberg,
Physica {\bf 117}A, 670 (1983).

\item{[25]} D. ben-Avraham,
{\sl Boundary and Finite-Size Effects in Lattice Models for
Dynamical Phase Transition}, preprint (1993).

\item{[26]} V. Privman,
Phys. Rev. Lett. {\bf 69}, 3686 (1992).

\item{[27]} S.N. Majumdar and C. Sire,
{\sl Phase Separation Model with Conserved Order Parameter
on the Bethe Lattice}, preprint (1993).

\item{[28]} K. Kang and S. Redner,
Phys. Rev. A{\bf 32}, 435 (1985).

}

\bye